\documentclass[journal]{IEEEtran}
\ifCLASSINFOpdf
  \usepackage[pdftex]{graphicx}
\else
\fi
\usepackage[colorlinks=true, citecolor=blue]{hyperref}

\begin{document}
%
\title{Generative AI and the Transformation of Software Development Practices}
%
%
%

\author{Vivek~Acharya {}
}

%
%

\markboth{Journal of \LaTeX\ Class Files,~Vol.~1, No.~1, May~2025}%
{Acharya \MakeLowercase{\textit{et al.}}: Generative AI and the Transformation of Software Development Practices}
%



\maketitle

\begin{abstract}
Generative AI is reshaping how software is designed, written, and maintained. Recent advances in large language models (LLMs) have enabled new development paradigms – from Chat-Oriented Programming (CHOP) and “vibe coding” to agentic programming – that promise accelerated productivity and expanded accessibility. This paper examines how AI-assisted techniques are transforming software engineering practices, alongside the emerging challenges of trust, accountability, and required skill shifts. We survey key concepts such as iterative chat-based development, multi-agent systems (agent clusters), dynamic prompt orchestration, and Model Context Protocol (MCP) integration. Through case studies and industry data, we illustrate both the opportunities (e.g., faster development cycles, democratized coding) and the complexities (e.g., model reliability, economic costs) of generative AI in coding. Our analysis provides a comprehensive overview of the ongoing generational shift in software development, delineating new roles, skills, and best practices for harnessing AI in an effective and responsible manner.
\end{abstract}


%
\IEEEpeerreviewmaketitle

\section{Introduction}
%
%
%
%
\IEEEPARstart{T}{he} rise of generative AI is driving a profound transformation in software development practices. Since the public debut of ChatGPT in late 2022, AI assistants based on large language models have rapidly gained adoption among programmers. By the end of 2023, an estimated 75\% of developers were using some form of AI-based coding tool in their workflow \cite{pragmaticengineer2025ai}. These tools – ranging from autocompletion engines to conversational coding assistants – leverage powerful LLMs to produce code, explain algorithms, and even generate entire software components from natural language prompts. As a result, tasks that once consumed hours of manual effort can now be completed significantly faster. A recent empirical study by McKinsey found that developers can finish certain coding activities in nearly half the time using generative AI assistance (e.g., writing new code ~50\% faster, and code refactoring ~33\% faster)\cite{karaci2025unleashing}. Such productivity gains hint at an impending leap in software engineering efficiency beyond what previous automation advances have achieved. However, the integration of AI into coding is not merely a quantitative acceleration of existing workflows – it is qualitatively changing how software is created. New development paradigms and terminologies have emerged to describe the evolving role of AI as a partner in the programming process. Chat-Oriented Programming (CHOP), a term popularized by Yegge and others in 2024, refers to coding via iterative dialogue with an AI, instead of the traditional “line-by-line” manual coding approach \cite{sourcegraph2024chop}. Likewise, the notion of “vibe coding” has arisen to describe a style of development where the programmer guides the AI through high-level intents and feedback, essentially “coding by feel” without focusing on the low-level details\cite{fisher2025vibe}. In parallel, researchers and practitioners are exploring agentic programming – harnessing semi-autonomous AI agents that can plan, write, and adapt code with minimal human intervention\cite{gaerber2025future}. These concepts signal a paradigm shift: rather than directly writing source code, developers are increasingly orchestrating and supervising AI-driven processes that generate the code. With these opportunities come new challenges. As AI takes on a greater share of code generation, issues of trust and accountability in software development are front and center. AI models do not inherently understand correctness or ethical use of code, and they can produce insecure or erroneous outputs. Organizations must therefore establish practices to ensure AI-generated code is reliable, secure, and compliant – raising questions about how to attribute authorship and responsibility between humans and AI. Furthermore, the economics of AI in development cannot be ignored: while AI assistants can boost individual productivity, the computational cost of large models has strained budgets, with industry surveys reporting nearly 89\% increases in computing costs from 2023 to 2025 largely due to AI adoption\cite{ibm2025ai}. Many companies have already postponed AI initiatives due to cost concerns\cite{ibm2025ai}, underscoring the need for cost-effective strategies in deploying these tools at scale. This paper provides a comprehensive exploration of Generative AI and the Transformation of Software Development Practices. In the following sections, we delve into the emerging paradigms (CHOP, vibe coding, agentic programming), and technical enablers (Model Context Protocol, agent clusters, dynamic prompting) that are shaping the future of coding. We discuss how software engineering teams can maintain trust and accountability in an AI-assisted workflow, and examine the evolving skill set required of “AI-era” developers. The generational shift underway – as a new cohort of AI-native developers enters the field and veteran engineers adapt – is analyzed, along with the broader economic and workforce impacts. Throughout, we cite recent case studies, industry benchmarks, and expert commentaries to illustrate these trends and best practices. By synthesizing insights across technical and human dimensions, our aim is to inform both practitioners and researchers about the current state and future trajectory of AI-powered software development.
\section{Chat-Oriented Programming (CHOP)}
One of the most prominent new paradigms enabled by generative AI is Chat-Oriented Programming (CHOP). Coined by Steve Yegge in mid-2024\cite{yegge2025death}, CHOP refers to a style of programming in which developers engage in an interactive conversation with an AI assistant to produce code, rather than writing code manually in a text editor. In essence, the coding process becomes dialogue-driven: the programmer specifies requirements, asks questions, and iteratively refines the output through prompts and natural language instructions, while the AI generates and modifies the code accordingly. This approach can be characterized as “coding via iterative prompt refinement”, as opposed to the classic approach of crafting each line of code by hand\cite{yegge2024chop}. 

CHOP in Practice: In a CHOP workflow, many traditional development steps are accelerated or altered by the AI’s capabilities. For example, understanding a new codebase – a task that normally involves reading documentation and source code for hours – can be streamlined by simply asking the AI pointed questions about the code. Developers report that with a chat-based assistant integrated into the IDE, “understanding our codebase becomes as simple as asking questions”, much like one would ask a teammate\cite{yegge2024chop}. The AI can leverage stored context or search the repository to explain functions, data models, and dependencies in seconds. This contrasts starkly with the pre-CHOP era, where a programmer might spend considerable time searching through code and external resources to build a mental model\cite{yegge2024chop}. When it comes to implementing new features or fixes, CHOP transforms the blank-screen problem into a guided co-creation process. Instead of manually writing boilerplate or looking up API usages, a developer using CHOP might “commission a painting” by describing the desired functionality in natural language\cite{yegge2024chop}. The AI then drafts code fulfilling that request, potentially offering multiple suggestions or improvements. The human can review this draft, ask the AI to adjust certain parts (e.g. “make this function iterative instead of recursive”), add constraints (“ensure the output is sorted”), or request tests. This interactive loop continues until the code meets the requirements. In effect, coding begins to resemble a high-level design discussion, where the developer focuses on what the program should do, and the AI handles much of the how. As one practitioner describes, “with CHOP, writing code is more like having a collaborative partner – you state what you want, and the AI fills in the details”\cite{yegge2024chop}. 

Impact on Development Workflow: CHOP significantly alters the software development lifecycle. Ado Kukic notes that in a traditional workflow, only one out of six typical development steps (coding) occurs inside the code editor, whereas tasks like requirement analysis, research, and code review occupy the majority of a developer’s time\cite{yegge2024chop}. Chat-oriented programming brings more of these activities into the conversational interface. For instance, researching solutions can be done by asking the AI for relevant algorithms or library suggestions, effectively offloading some Google searches to the assistant (which may have been trained on extensive documentation). Even code review can become a collaborative process with the AI – developers can ask the assistant to explain a code change or check it against certain style guidelines before raising a human code review. 

Early evidence suggests CHOP can accelerate development without sacrificing quality, provided developers maintain oversight. Test-Driven Development (TDD) can mesh well with CHOP: the developer specifies the tests (in natural language or code), and the AI generates code to pass those tests\cite{yegge2024chop}. The tests act as a correctness contract, helping ensure the AI’s output meets the specifications. This is one example of how trust is managed in CHOP – by anchoring AI outputs to verifiable criteria. In Section VI, we will discuss trust and accountability measures in more depth.

CHOP vs. Traditional Prompt Engineering: It is important to distinguish CHOP from the simpler notion of one-shot prompt engineering. In the early days of coding with GPT-3/ChatGPT, developers would often write a single elaborate prompt (i.e., a detailed instruction describing the entire task) and hope the model’s single response solved it. CHOP instead embraces an interactive, iterative approach – the prompt is refined through conversation. Rather than spending excessive effort upfront to engineer the perfect prompt, the developer guides the AI step by step, which is more forgiving and adaptive. This dynamic aligns with modern dynamic prompting techniques (Section IX), which leverage ongoing context and dialogue, instead of static prompts. In summary, CHOP is programming with conversation: the code emerges through back-and-forth communication between human and AI, leading to a more fluid and potentially more creative development process.
\section{Vibe Coding}
Another novel concept gaining attention in the developer community is “vibe coding.” Introduced by AI researcher Andrej Karpathy in early 2025, vibe coding refers to a highly intuitive way of coding with AI where the developer “fully gives in to the vibes” of the AI’s suggestions\cite{willison2025vibe}. In practical terms, vibe coding means expressing one’s intent or vision for an application in natural, perhaps even colloquial language, and allowing the AI to generate and modify the code with minimal manual intervention or meticulous scrutiny from the developer. It is a radically hands-off approach: the programmer focuses on guiding the feeling or high-level behavior of the program, while trusting the AI to handle the low-level implementation details. 

Definition and Philosophy: As described by Karpathy, vibe coding involves “forgetting that the code even exists” and instead interacting with the development environment almost as if one were chatting with a human collaborator\cite{willison2025vibe}. The developer might say things like, “Make the sidebar’s padding half as large” or “The animation should feel more responsive to the music beat”, and the AI would apply these changes to the codebase. This paradigm is enabled by increasingly powerful AI coding assistants (e.g., enhanced IDEs like Cursor with voice control) that can interpret such natural-language directives and map them to code edits. Karpathy even demonstrated using voice commands (via OpenAI’s Whisper for speech-to-text, dubbed “SuperWhisper”) to converse with the IDE, thereby hardly touching the keyboard\cite{willison2025vibe}. In vibe coding, the developer often accepts the AI’s outputs with little to no manual adjustment – a stark contrast to normal practice where code reviews and careful reading are mandatory. 

Advantages and Use Cases: The primary appeal of vibe coding is speed and creative flow. By minimizing context switches (the programmer doesn’t have to search files or write boilerplate) and by leveraging the AI’s ability to generate code rapidly, prototypes can be built extremely quickly. One developer reported being able to get a weekend project web app “up and running in a single sitting” by leaning on vibe coding, whereas manually coding it might have taken days\cite{fisher2025vibe}. The “code first, refine later” mindset behind vibe coding encourages experimentation: developers can try wild ideas or implement features on a whim, knowing that the AI can produce a baseline implementation swiftly\cite{ibm2025vibe}. This approach aligns with agile prototyping – get something working (even if imperfect) and iterate. It also lowers the barrier for non-experts to create software; someone with an idea but limited programming knowledge can describe the idea in plain language and let the AI do the heavy lifting of actual coding\cite{ibm2025vibe}. In fact, vibe coding is often touted as making the act of coding more accessible and low-friction (see Section XIV), potentially enabling domain experts or designers to build software by describing what they want. 

Risks and “House of Cards” Code: Despite its allure, vibe coding comes with significant caveats. By design, this style eschews the rigorous critical evaluation of AI-generated code. Simon Willison remarks that “vibe coding is not the same as responsible AI-assisted programming”\cite{willison2025vibe}. Professional developers have an obligation to ensure code is correct, maintainable, and efficient – tasks that require reading and understanding the code. In vibe coding, the developer might accept AI outputs without closely examining them, leading to what Addy Osmani calls “house of cards code” – it appears to work but can collapse under real-world conditions\cite{osmani2025ai}. For example, an AI might produce a function that works on simple test cases but fails on edge cases or has security flaws. If the developer never inspects this function (trusting the vibe), these issues remain hidden until they cause a failure in production. Thus, while vibe coding can be “amusing” and productive for low-stakes projects\cite{willison2025vibe}, it is generally ill-suited for critical production software unless paired with thorough validation steps. 

Responsible Use of Vibe Coding: The community consensus is that vibe coding should be confined to scenarios like rapid prototyping, hackathons, or exploratory programming where speed is valued over robustness. Even then, developers must eventually switch back to a more traditional mode to harden the code: review it, add tests, refactor messy AI-generated structures, and ensure the “vibe” implementation meets non-functional requirements (performance, security, etc.)\cite{willison2025vibe}. One can think of vibe coding as an extreme point on the spectrum of AI assistance – maximizing convenience and relying on AI for virtually everything – whereas in professional settings a balance must be struck. The AI can draft large swaths of code at a high level, but a human should verify and integrate it conscientiously. As Willison notes, “my golden rule for production-quality AI-assisted programming is that I won’t commit any code I couldn’t explain”. This highlights that even if vibe coding gets something working, engineers should personally understand the codebase before shipping it. In summary, vibe coding exemplifies the new possibilities (and perils) when development becomes more about guiding an AI’s “intent” rather than writing code – it’s a powerful accelerant for creativity, best used with caution and followed by disciplined engineering practices.
\section{Agentic Programming}
While CHOP and vibe coding focus on interactive human-AI collaboration at development time, agentic programming extends the role of AI into more autonomous, run-time decision-making and coding tasks. Agentic programming is an emerging paradigm where developers build systems by deploying AI agents that can perform complex tasks independently – such as writing code, fixing bugs, or optimizing performance – by reasoning about goals and taking actions without step-by-step human guidance. This approach represents a shift from writing explicit instructions (code) to designing intelligent systems that figure out the instructions on their own\cite{gaerber2025future}. In essence, instead of manually coding every behavior, a developer orchestrates a team of AI agents, each with certain capabilities, and the agents collaborate to achieve the developer’s objectives. Definition and Rationale: An AI agent in this context typically refers to an LLM-based process augmented with tools, memory, and sometimes feedback mechanisms. Unlike a single LLM prompt-response, an agent maintains state over multiple steps, can call external APIs or functions, and can decide on its next action based on intermediate results (this is often implemented via a reasoning loop or frameworks like ReAct). Agentic programming uses such agents as the building blocks of software. Instead of writing code to solve a problem, the developer specifies the problem to a set of agents and provides them with the means to solve it (access to data, tools, and perhaps a high-level strategy). The agents then autonomously break down the problem, coordinate among themselves, and produce a solution – which could include generating code, executing it, testing outcomes, and revising as needed\cite{gaerber2025future}. 

For example, imagine a scenario where a “Bug-Fixer” agent monitors a code repository. When it detects a failing test, it analyzes the error, localizes the offending code, and attempts a fix by modifying the code. It could use a “Compiler/Runner” agent to run the tests after the fix. If tests still fail, it iterates, possibly consulting a “Research” agent to search documentation for error messages. This kind of setup moves us towards self-healing software and automated maintenance – a core promise of agentic programming. 

Key Components and Frameworks: Implementing agentic programming requires several ingredients: (1) Planning – the ability for agents to break tasks into subtasks and decide on actions (often using an LLM’s chain-of-thought capabilities); (2) Memory/State – mechanisms for agents to remember context beyond a single prompt (via embedding vectors, scratchpad memory, etc.); (3) Tool use – enabling agents to call external tools (compilers, web search, database queries, etc.) to act on the world; and (4) Orchestration – a way to have multiple agents communicate or a controller to manage the agents. Research and open-source projects have proliferated in this area. For instance, Microsoft’s Autogen framework provides “conversable and customizable agents” with features like nested chat (hierarchical tasks) and group chat (multiple agents solving tasks together)\cite{pan2025agentic}. These frameworks exemplify multi-agent orchestration, where agents might play different roles (e.g., a “Planner” agent and a “Solver” agent, or a group of peer agents each tackling part of a problem)\cite{pan2025agentic}. 

A salient aspect of agentic systems is that performance can scale with model improvements. Because agents rely on underlying LLMs for reasoning, a more powerful model can make an existing agent system immediately more capable\cite{gaerber2025future}. This creates a compelling feedback loop in software development: rather than rewriting code to improve an algorithm, one could upgrade the AI model that agents use, and they will autonomously handle more complex tasks or handle simple tasks more reliably. 

Applications and Early Results: Agentic programming is still nascent, but early use cases show promise. Complex workflows that are cumbersome to hard-code can be naturally handled by agents. For example, an AI code assistant agent might be tasked with implementing a feature: it can generate code, run tests to validate it, detect failures, and adjust code in a loop until tests pass – essentially performing a rudimentary form of development and debugging by itself. In one experiment, researchers found that an agent-based approach could successfully fix vulnerabilities in code by iteratively testing and patching, demonstrating a kind of automated debugging skill\cite{forte2025multiagent} \cite{automq2025wiki}. In production settings, companies are exploring agents for tasks like customer service bots that troubleshoot technical issues or DevOps agents that monitor systems and execute routine remediation steps. 

However, current agentic systems also face challenges. One is reliability – agents can get stuck in loops or pursue wrong solutions if their reasoning drifts. Ensuring they know when to stop or when to ask for human help is an open problem (the so-called “AI alignment” at a micro level). Another challenge is providing the right environment: an agent needs a lot of context (codebase, documentation) and the right tools to be effective. This is where the Model Context Protocol (MCP) comes in (see Section VII), providing standardized access to data and tools for agents. In summary, agentic programming shifts the developer’s role from writing detailed algorithms to defining goals and equipping AI agents to achieve those goals. It holds the potential to automate not just coding, but higher-level software engineering activities (like architecture search, optimization, and maintenance). As agent frameworks and techniques mature, we may see software projects where human engineers act more as product managers or supervisors to fleets of coding agents – setting objectives, constraints, and reviewing outcomes, rather than crafting every line of code themselves. The next sections discuss two important concepts related to agentic programming: the Model Context Protocol which enables agents to interface with real-world data, and agent clusters (or swarms) which involve multiple agents collaborating on tasks.
\section{Trust \& Accountability}
As generative AI becomes deeply integrated into software development, maintaining trust in the code and establishing clear accountability for its quality is paramount. In traditional development, trust is built through code reviews, testing, and the understanding that a known developer wrote the code and stands behind it. Generative AI disrupts this model: when an AI assistant writes a portion of code, developers might be tempted to accept it without full comprehension, and it may be unclear who “owns” the correctness of that code. Organizations adopting AI therefore face a “trust gap” – how to ensure AI contributions meet the same standards as human code, and how to attribute responsibility for those contributions.

Erosion of Implicit Trust: In a conventional workflow, there is an implicit chain of trust. Engineers adapt code from known sources or write it themselves, understanding its logic in the process, which creates a sense of ownership\cite{sonarsource2025trust}. With AI-generated code, this chain can break. An AI coding assistant can produce a snippet that fits seamlessly into the project context (thanks to training on vast code corpora and awareness of local variables), meaning the developer might need only minimal tweaks to integrate it. On one hand, this is efficient; on the other, it can lead to “blind acceptance” of code that the developer has not mentally executed or rigorously analyzed. The result is code in the repository whose true author is an AI (even if a human pressed enter) – raising questions: Can the team trust this code? Who is accountable if it fails or has bugs?

Moreover, the ease of obtaining code from AI can encourage bypassing best practices. If an organization has no guidelines on AI usage, developers might use public AI services in “shadow IT” fashion. This could introduce licensed code unknowingly or leak proprietary logic into external model queries\cite{sonarsource2025trust}. Without oversight, the provenance of code becomes murky – an auditor might find it hard to tell which code was human-written and which came from an AI (and if so, which AI and with what prompt).

Establishing Accountability Frameworks: To tackle these issues, leading companies and tool providers are developing frameworks to embed accountability and transparency into AI-assisted development. These involve process changes and tool support to track AI contributions. For example, SonarSource suggests tagging any code contribution that was AI-generated in the version control system\cite{sonarsource2025trust}. If a developer uses an AI assistant to create a new function, the commit could be labeled (by the IDE or a hook) as “AI-assisted” along with metadata such as the model name and version used. This creates an audit trail: later, if that code is found defective, engineering managers can trace it back to an AI generation event, and treat it akin to code written by a junior developer – i.e., recognize it may need closer review. Figure 1 illustrates a recommended process for integrating AI into development with governance steps to maintain trust.
\begin{figure}
    \centering
    \includegraphics[width=1\linewidth]{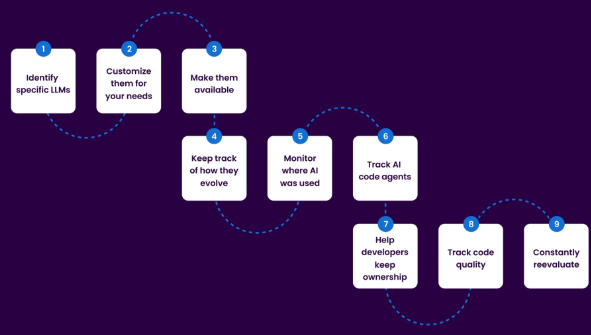}
    \caption{Integrating AI into software development with trust and accountability measures}
\end{figure}

Fig. 1. Integrating AI into software development with trust and accountability measures. A stepwise framework (1–9) emphasizes selecting approved models, customizing and monitoring them, tagging AI-generated code for transparency, and keeping developers in the loop for validation

Another crucial aspect is restricting AI usage to vetted models and tools. Organizations can approve specific LLMs (or run their own) that developers are allowed to use, considering factors like security and code licensing\cite{sonarsource2025trust}. By doing so, they reduce the risk of introducing unknown liabilities (for instance, an open-source model trained on copyleft code might regurgitate licensed snippets). Some companies fine-tune these models on their proprietary codebase to better align outputs and then require all AI suggestions to come from this in-house model. This ensures the AI is domain-aware and also that any data given to the model stays internal.

Perhaps the most important element in maintaining accountability is keeping the developer in the loop as the final arbiter. Teams are instituting policies that even if AI writes code, a human developer must review and take ownership of it before it’s merged. In other words, the AI can propose, but a human disposes (approves). Some development teams treat the AI like an “intern” or “very eager junior developer” – capable but needing oversight\cite{osmani2025ai}. The human reviewer ensures that any AI-produced code segment is well-understood, meets style and quality norms, and is accompanied by tests. Tools can assist here as well: AI-generated code can be automatically run through static analysis and testing pipelines (e.g., using SonarQube or similar) to catch common issues, giving the human reviewer more confidence in its correctness.

Accountability Models: On a governance level, companies are exploring models of accountability for AI contributions. One approach is to extend code review checklists to include AI-specific items: for example, reviewers might ask, “Was this code AI-generated? If so, did we verify that it doesn’t duplicate licensed code and that the logic is sound?” Another approach is documentation: requiring developers to document in commit messages or design docs when AI was used, and any additional assumptions or prompt instructions given. This creates a knowledge base for future maintainers. 

In regulated industries or safety-critical software, stricter accountability might be necessary. We could envision a future where certain code must be “AI-free” (fully written and verified by humans) or where AI usage needs sign-off from a project lead. Conversely, for non-critical code (like internal tools or prototypes), teams might allow more liberal use of AI and tolerate the minor risk in exchange for higher velocity.

In conclusion, trust and accountability in an AI-assisted development environment require a combination of cultural practices and technical solutions. Culturally, developers should treat AI as a powerful assistant that still requires their guidance and verification – much like pair programming with a junior colleague. Technically, organizations should track AI involvement and enforce checkpoints (like tagging, testing, and approval gates) to ensure that the convenience of AI does not erode the quality and reliability of the software. When implemented well, these measures allow teams to enjoy the productivity benefits of generative AI without compromising on trust in the final product.

\section{Model Context Protocol (MCP)}
As developers embrace chat-based and agentic programming, a critical technical challenge arises: how to feed the AI models with the right context and data from the software project (and its environment) so that the AI’s outputs are accurate and relevant. Large language models in isolation are “stateless” beyond their prompt – they do not inherently know about your specific codebase, requirements, or runtime environment unless that information is provided in context. Early approaches to this problem involved ad-hoc solutions like copy-pasting relevant code into the prompt or building custom connectors for each data source an AI might need (e.g., one for documentation, one for a database). This approach does not scale well. The Model Context Protocol (MCP) emerged in late 2024 as an open standard to tackle this problem by creating a uniform way for AI assistants to interface with external data, tools, and contexts\cite{anthropic2025modelcontext}.

Overview of MCP: The Model Context Protocol is essentially a “USB-C for AI” – a standardized port through which AI models (clients) can securely connect to various data sources and services (servers)
anthropic.com
. Announced by Anthropic and partners, MCP defines how an AI agent can request data or actions from a context server, and how that server should respond. For example, a context server could expose a project’s Git repository or documentation. An AI coding assistant that implements MCP could then query, say, “get file content of utils.py” or “search for function X in the repository” via the protocol, rather than relying on the prompt to contain that info from the start. The goal is to move away from isolated, siloed AI usage (where every new integration requires custom code) towards a plug-and-play ecosystem\cite{anthropic2025modelcontext}. Ecosystems such as iHE rely on governed data exchange (e.g., FHIR/TEFCA); mapping these into MCP would let agents consume clinical context safely and reproducibly \cite{acharya2025intelligent}.

In architecture terms, developers can run lightweight MCP servers that wrap around specific resources: one might wrap a codebase and allow queries like file retrieval or code search; another might wrap a bug tracking system to fetch issue details; another could provide database queries or live system metrics. On the other side, an AI application (like an IDE plugin or an autonomous agent) acts as an MCP client, requesting and receiving context through a standardized API. 

\textbf{Benefits for Software Development}: By using MCP, AI assistants can maintain a much richer and up-to-date context about the project they are working on. Instead of hitting token limits by stuffing the entire relevant code into a prompt, the assistant can call MCP endpoints as needed to fetch only what is necessary. This not only improves the quality of responses (the AI always has access to the latest data) but also enhances security and auditability – since all data access is via defined channels, it’s easier to log and control what the AI saw. For example, an MCP server for source control could enforce permissions, ensuring the AI only reads files the developer is allowed to see, or scrub sensitive information before returning results\cite{descope2025mcp}.

Anthropic reported that companies like Block (Square) and development platforms such as Replit, Sourcegraph, and Zed began collaborating on MCP, seeing it as an enabler for more powerful coding AI integrations. These platforms found that MCP can help AI agents “retrieve relevant information to understand the context around a coding task and produce more nuanced code with fewer attempts”\cite{anthropic2025modelcontext}. In practical terms, this means fewer hallucinated answers or irrelevant suggestions – the AI can directly ask for context it doesn’t know, rather than guessing.

\textbf{Example Use Case}: Consider a scenario of CHOP where a developer asks, “AI, how is the user’s session managed in this application?” Without MCP, the AI might try to recall from training or require the developer to paste code. With MCP, the AI can query the MCP server connected to the codebase: perhaps it issues a search query for “session management” and finds that the app uses, say, a Redis store and a specific utility function. The MCP server returns the snippet or description, and the AI can now give an accurate answer, citing the actual code. Similarly, an agent trying to debug could fetch the last 100 lines of a log file via MCP to see what error occurred, then proceed accordingly.

\textbf{MCP and Agentic Systems}: For agentic programming (Section IV), MCP is a game-changer. Autonomous agents need reliable ways to get information about their environment. By programming an agent to use MCP calls, we give it structured tools to gather data and perform actions safely. In essence, MCP can serve as the “eyes and hands” of an AI agent in a software system, where the core LLM provides the “brain.” This separation of concerns (brain vs. I/O interface) simplifies the design of agents. It also improves security: since MCP can enforce that certain destructive actions (like writing to a database) require explicit permissions or confirmations, it mitigates the risk of an agent going rogue or making unintended changes.

\textbf{Standardization and Adoption}: MCP is an open protocol, and its adoption is growing. Having an industry standard means tools from different vendors can interoperate. An AI assistant from Company A could connect to a context server by Company B as long as both speak MCP. This fosters an ecosystem where best-of-breed context providers (for code, for documentation, etc.) can be mixed and matched with various AI models and agents. Over time, this could lead to a robust marketplace of MCP-compatible tools: e.g., specialized MCP servers for popular frameworks (Django, React, etc., providing context about framework internals), or for DevOps systems (Kubernetes control, cloud monitoring data, etc.).

From a developer’s perspective, MCP’s promise is that your AI buddy will always be aware of the key context – your code, your data, your tools – without you having to manually feed it. As one CTO put it, “Open technologies like the Model Context Protocol are the bridges that connect AI to real-world applications… removing the burden of the mechanical so people can focus on the creative”\cite{anthropic2025modelcontext}. In summary, MCP is a foundational piece for the next generation of AI-assisted development, ensuring that AI tools are context-aware, secure, and interoperable. It shifts the focus from wrangling prompt data to actually solving problems, by letting the AI access what it needs when it needs it, much like a developer would consult documentation or inspect code while coding.

\section{Agent Clusters}
Building on agentic programming and protocols like MCP, developers are now experimenting with orchestrating multiple AI agents working in concert, often referred to as agent clusters or AI swarms. An agent cluster is a collection of AI agents, each potentially with specialized roles or expertise, that collaborate towards a common objective. This concept extends the single-agent paradigm to a multi-agent system where coordination, communication, and emergent problem-solving become possible. In many ways, this mirrors how complex software today is built by teams of specialists (frontend, backend, QA, etc.), but here the “team members” are AI agents.

Concept and Motivation: No single agent, no matter how advanced, is universally best at all tasks. By clustering agents, we can leverage their diverse strengths. For example, one agent might be optimized for code generation, another for testing and verification, and another for planning and decomposition of tasks. Working together, they can achieve outcomes that are difficult for a lone agent. OpenAI’s research hints at this direction by enabling agents that can handle multi-stage reasoning, replan on the fly, and effectively “reason together” rather than following a static script\cite{codewave2025swarms}. The practical implication is significant: “multi-agent coordination is no longer theoretical”\cite{codewave2025swarms}, meaning real software can be built or managed by a fleet of agents that dynamically assign tasks among themselves.

\textbf{Examples of Agent Clusters}:

\begin{itemize}
    \item Planning and Execution Cluster: A simple cluster might involve a Planner agent and several Worker agents. The Planner breaks a software task (say, “implement a feature X”) into sub-tasks and assigns them to Worker agents. Each Worker (perhaps an instance of an LLM with coding capability) tackles its sub-task (one might handle database changes, another the UI changes, etc.). The Planner then integrates their outputs. This divide-and-conquer strategy can parallelize development and play to each agent’s strengths, and is supported by frameworks like AutoGen which explicitly provide “Group Chat” for multiple agents to converse and solve tasks collectively\cite{pan2025agentic}.
\end{itemize}
\begin{itemize}
    \item Competition and Refinement Cluster: Another pattern is having multiple agents attempt the same task in parallel (or alternate) and then compare results. For example, two code-generation agents could produce solutions\cite{yegge2025death} for a bug fix; a third agent (judge) compares the outputs (or runs tests) and selects the better solution.This mimics code review or pair programming, introducing redundancy to increase reliability.
\end{itemize}
\begin{itemize}
    \item Specialist Cluster: We might also see clusters where each agent has access to different tools or knowledge. One agent might have access to a database (for data analysis tasks), another to the internet (for knowledge lookup), and another to the codebase. Working together, they handle tasks requiring cross-domain knowledge – similar to how microservices communicate, but here at an AI agent level.
\end{itemize}

\textbf{Coordination Mechanisms}: In an agent cluster, communication is key. Agents may communicate via natural language messages (as if in a chat room) or via a shared memory/state. Techniques from distributed systems apply: establishing protocols for consensus, handling conflicting actions, etc. A notable approach in research is to use a central cognitive architecture or a shared workspace where agents post intermediate results and observations (a blackboard system). Alternatively, completely decentralized swarms have been proposed, where agents self-organize. Amazon, for example, demonstrated a multi-agent collaboration system for complex state management, effectively showing how to maintain consistency across agent clusters that share partial information\cite{galileo2025agentic}.

\textbf{Benefits}: Agent clusters can tackle complex, multifaceted problems more effectively. A use case described in recent literature is an “Autonomous Research Assistant” where a cluster of agents handles different parts of research – one reads papers, another extracts key points, another checks consistency – and collectively they produce an analysis\cite{codewave2025swarms}. The cluster can cover more ground and cross-verify findings, akin to a team of researchers. In software engineering, imagine a cluster managing an entire project: one agent writes code, another continuously runs the test suite, another monitors performance metrics and suggests optimizations, all in a loop. Such a cluster could, in theory, iterate towards a working, tuned application with minimal human oversight.

\textbf{Challenges}: With multiple agents, emergent behavior can be both a feature and a bug. Coordination breakdowns can happen – e.g., agents might get stuck passing tasks back and forth (“you do it” – “no, you do it”), or they might overload a system with too many simultaneous changes. Ensuring that the cluster’s behavior is stable and predictable is non-trivial. Researchers often constrain interactions to avoid infinite loops or require a human to serve as a final arbiter if agents disagree irreconcilably. Performance is another factor: running many large models in parallel is resource-intensive, though some tasks might not need all agents active at once.

\textbf{Agent Clusters vs. Traditional Concurrency}: It’s worth noting that multi-agent AI systems have analogies to concurrent programming. Concepts like deadlock, starvation, and race conditions might well apply when agents operate on shared resources (like editing the same code). Solutions from that domain – locks, transaction protocols – might inform multi-agent system design. The difference is that agents, powered by AI, have more flexibility and unpredictability than hard-coded threads. They can negotiate, learn from each other’s output, and even reconfigure roles on the fly. For instance, one agent might detect that another agent’s approach is failing and decide to take over that task or spawn a new helper agent. This adaptability is powerful, but also hard to formally verify for correctness.

Current State and Outlook: At present, agent clusters are mostly in experimental and early-adopter stages. We see them in sophisticated applications like cybersecurity defense, where multiple agents watch different parts of a network and collaboratively respond to threats in real-time\cite{codewave2025swarms}. In that scenario, a swarm of agents can cover a broad attack surface and react faster than a centralized system, since each agent is semi-autonomous and specialized. In software development, fully autonomous swarms delivering end-to-end projects are not mainstream yet, but prototypes exist (for example, AutoGPT experiments that try to build simple apps with a team of AI agents). Over the next few years, we can expect better frameworks to manage agent clusters, borrowing from both distributed AI research and practical lessons of deploying these systems. If successful, agent clusters could become a new unit of computing – just as we scale software by adding more servers or microservices, we might scale development or maintenance by adding more AI agents to the cluster.

In conclusion, \textbf{agent clusters} represent the collaborative dimension of AI in software engineering. They extend the capability of individual agents by enabling teamwork, specialization, and parallelism. For developers and engineering managers, learning to design and steer these clusters will be an important skill – akin to managing a highly automated dev team. With proper coordination and oversight, agent clusters (or AI swarms) could tackle large-scale software challenges, maintain complex systems in production, and accelerate innovation in ways that even a skilled human team might struggle to match, all while operating under human-defined goals and constraints.

\section{Prompt Engineering vs Dunamic Prompting}
Throughout the evolution of AI-assisted development, one skill that has often been highlighted is prompt engineering – the craft of writing effective inputs to guide LLMs toward desired outputs. However, as our use of AI grows more sophisticated, a shift is occurring from static prompt engineering to dynamic prompting strategies. This section contrasts the two approaches and explains why dynamic prompting is becoming crucial in modern AI-enhanced software workflows. 

\textbf{Traditional Prompt Engineering}: In the earlier days of using models like GPT-3 for coding (2020–2022), developers learned that the phrasing and details of their prompts dramatically affected the results. Prompt engineering involved carefully designing the input query: for example, telling the model its role (“You are an expert Python developer…”), specifying the format of the output (e.g., “provide only the code without explanations”), and even giving examples in the prompt. This was often a manual, trial-and-error process where one aimed to find a “magic incantation” that yielded correct and coherent results. In a coding context, a prompt might explicitly include relevant API documentation or a partially written function to nudge the model in the right direction. Prompt engineers collected patterns (like including edge case descriptions or using particular keywords to trigger certain behaviors) and these became a kind of folk knowledge for interacting with LLMs.

While prompt engineering is powerful, it treats the model essentially like a black box that one has to prompt perfectly in one go. The prompt is static text, and if the output is not correct, the main recourse is to rewrite the prompt and try again (or tweak a few parameters). This approach can hit limitations: for complex tasks, a single prompt might not suffice to encode all nuances; and long prompts can exhaust token limits or confuse the model if not structured well.

\textbf{Dynamic Prompting}: In contrast, dynamic prompting is an approach where the interaction with the model is adaptive, multi-turn, and contextually evolving\cite{q3tech2025dynamicprompts}. Instead of a one-shot prompt, the developer (or an agent) builds the prompt context dynamically, incorporating intermediate results, user inputs, or external data in real-time. Dynamic prompting lies at the heart of advanced LLM applications and frameworks today. Some key characteristics include:

\begin{itemize}
    \item \textbf{Adaptive Context}: The prompt to the model can change based on what the model previously output or what the user has done. For instance, if an AI agent trying to write code encounters a compilation error, it will take that error message and dynamically insert it into the next prompt, asking the model to fix the issue. This is a form of error-driven prompting – not predetermined, but generated in response to events.
\end{itemize}
\begin{itemize}
    \item \textbf{Personalization and Memory}: Dynamic prompting enables maintaining a memory of the conversation or session. The system might prepend a summary of past interactions or user preferences to the prompt. This allows a more personalized and coherent long-running dialogue, beyond the fixed-length context window. For example, a chatbot coding assistant could remember which frameworks the user prefers and dynamically bias future prompts toward those, yielding more relevant suggestions.
\end{itemize}
\begin{itemize}
    \item \textbf{Programmatic Prompt Construction}: Tools like LangChain provide mechanisms to construct prompts from multiple pieces: e.g., an initial system prompt that sets the stage (perhaps the role and persona of the AI), followed by user query, followed by relevant context fetched from documents (using retrieval), etc. This assembly happens behind the scenes for each query, based on pipelines or chains the developer sets up. It’s “prompt engineering” in a broader sense, but much of it is automated and reactive. For instance, a chain might automatically perform a keyword search in documentation and include the top result in the prompt when the user asks a question about a library function.
\end{itemize}
\begin{itemize}
    \item \textbf{Multi-turn Orchestration}: In dynamic prompting, one prompt’s output can directly lead into the next prompt as input. This effectively creates an implicit loop or conversation. Consider a scenario: The user says, “Generate a function for X.” The AI returns code. The system then automatically prompts, “Now generate tests for that function” (without the user explicitly asking), using the previous answer as context. This chain of prompts is dynamic – it wasn’t all written upfront by the user, but orchestrated by a higher-level script or agent logic. It results in a more interactive and thorough experience (in this example, ensuring that whenever code is generated, tests follow).
\end{itemize}

The benefits of dynamic prompting are clear in terms of contextual awareness and tailored responses. A static prompt might produce a generic answer, but a dynamic system can make the AI aware of the current context, such as the exact error the code produced, or the part of the codebase the user is working in\cite{q3tech2025dynamicprompts}. This leads to responses that are more accurate and useful. In fact, dynamic prompting is largely responsible for making AI assistants practically viable – it’s what allows ChatGPT or Copilot to feel “interactive” and not just like a Q\&A bot.

\textbf{Example – Bug Fixing Loop}: To illustrate, suppose a developer is using an AI agent to resolve a bug. Using dynamic prompting:
\begin{enumerate}
    \item The agent asks the model: “Given the code snippet (with bug) and the failing test output, suggest a fix.” (This prompt is constructed with the current code and error message included.)
    \item The model suggests a code change.
    \item The agent applies the change and runs tests, finding a new error.
    \item The agent then dynamically prompts: “The fix was applied, but now we see this new error (include error text). How to address that?”
    \item The model replies with another adjustment.
    \item This loop continues until tests pass.
\end{enumerate}

In this process, each prompt after the first is generated dynamically based on the latest outcome (new error messages). Traditional prompt engineering alone could not achieve this – one would not be able to foresee all possible errors to pre-write in a single prompt.

\textbf{Prompt Engineering in the Era of Dynamic Prompting}: It’s worth noting that dynamic prompting doesn’t eliminate the need for prompt engineering; it changes its nature. Instead of crafting one perfect prompt, developers now design prompt strategies or templates. They need to think about how to frame each step of a conversation or chain. For instance, one might prompt the model to first output a plan before writing code (“First, list the steps you will take, then we’ll proceed step by step”). Such meta-prompts help manage the reasoning process (this intersects with research on chain-of-thought prompting, where the model is asked to reason stepwise).

In summary, static prompt engineering is like writing a single question on a card and handing it to a genie, hoping for the best answer; dynamic prompting is like having an ongoing interview or dialogue with the genie, where each question builds on the last. The latter is evidently more powerful for complex tasks akin to software development, which inherently is an iterative process.

The transition to dynamic prompting reflects a maturation in how we utilize AI: from a novelty where one needed clever tricks to coax good output, to a more systematized approach where AI interactions are built into the software development pipeline itself. As one industry report noted, “advanced prompt engineering techniques – including dynamic prompts – are at the forefront of AI evolution, enabling LLMs to deliver real-time contextual understanding like never before”\cite{q3tech2025dynamicprompts}. Developers who embrace dynamic prompting will be able to harness AI far more effectively than those sticking to static prompts, especially as projects scale in complexity.

\section{Skill Transition}
The advent of generative AI in software engineering is prompting a significant skill transition for developers and teams. As mundane coding tasks become automated or accelerated by AI, the human role is shifting towards higher-level decision making, oversight, and integration of AI tools into the development process. This section discusses how developers’ skillsets are evolving and what new competencies are becoming important. 

\textbf{From Coding to Curating}: Traditionally, a junior software engineer might spend much of their time writing boilerplate code, fixing simple bugs, or translating specifications into code – essentially learning by doing the grunt work. With AI assistants, many of these entry-level tasks can be done partially or wholly by the AI. This means junior developers might find themselves in a position of having to review and curate AI-generated code rather than write everything from scratch. On one hand, this can accelerate their exposure to more complex code (since the AI can generate relatively advanced patterns that they can learn from). On the other hand, it can be challenging because the traditional learning-by-doing path is disrupted. Experts caution that junior engineers often miss crucial steps when relying on AI output – they might accept the AI’s code without fully understanding it, leading to fragile “house of cards” solutions\cite{osmani2025ai}. To transition effectively, these engineers need to develop skills in critical evaluation of AI output: understanding what the code does, adding appropriate tests, and not being lulled into a false sense of security by seemingly working code.

\textbf{Prompting and Orchestration Skills}: As discussed in Section IX, being adept at formulating prompts and orchestrating AI interactions is now a valuable skill. Developers are learning to become good at describing problems to AI (a bit like how one learns to ask a question on Stack Overflow clearly). But beyond that, they are learning to design entire workflows around AI. This includes knowing when to break a problem into subprompts, how to use tools like retrieval augmentation (e.g., vector databases) to supply context to the model, and how to interpret partial results to guide the next prompt. In essence, developers are learning to program on a meta-level: programming the AI through instructions rather than writing program code directly. Some have dubbed this “AI orchestration” or “meta-coding”, and it involves frameworks (LangChain, etc.) as well as conceptual thinking about AI behavior. This is a new skill set that was not part of traditional software engineering education.

\textbf{AI-Augmented Design and Architecture}: With AI handling coding details, human developers can focus more on design, architecture, and requirements – the creative and analytical parts of engineering. We foresee skills like system design becoming even more important. A developer might sketch a high-level solution and then rely on AI to fill in the boilerplate in each component. The skill lies in getting the specifications right and ensuring that the overall architecture the AI is filling in is sound. Additionally, because AI can generate multiple alternatives quickly, engineers need decision-making skills to choose the best among the AI’s suggestions. For example, if an AI proposes three different code refactoring approaches, the engineer must evaluate trade-offs (performance, readability, etc.) to pick one. Thus, the ability to evaluate and compare solutions – a higher-order skill – becomes more prominent than the ability to craft any single solution manually.

\textbf{Continuous Learning and Adaptability}: The AI field is fast-moving. Models can change in capability with new versions (GPT-3 vs GPT-4 vs GPT-5, etc.), and new tools come out frequently. Developers are now expected to continuously learn not just new programming languages or libraries, but new AI features and paradigms. Those who adapt quickly gain an edge. For instance, when multi-modal models (that can see images or diagrams) became available, some developers leveraged them to debug UI issues or interpret graphical data. Teams that stay updated on AI capabilities can integrate them creatively (like using an AI agent to optimize database queries or to generate infrastructure-as-code scripts). This puts a premium on AI literacy as a core skill – understanding at a conceptual level how LLMs work, their limitations (e.g., knowledge cutoff, hallucination tendencies), and keeping abreast of enhancements.

\textbf{Roles and Mentorship}: We may also see the emergence of roles like AI Engineer or Prompt Engineer formally in teams, though these may be transitional titles. An AI Engineer would specialize in integrating AI into products – selecting models, fine-tuning them, building prompt pipelines, and so on. Already, job postings for prompt engineering and LLM integration have appeared in industry. For seasoned developers, mentoring juniors now includes teaching them how to use AI tools effectively and ethically. It’s less about teaching syntax or basic algorithms (the AI can help with that) and more about teaching critical thinking, debugging strategies, and when to trust or double-check the AI. The mentor’s role evolves to ensure that newcomers still learn fundamental concepts even if AI handles many details. One approach is to have juniors re-implement or at least simulate parts of what the AI did, to grasp the underlying principles.

\textbf{Mindset Shift}: There is also a psychological aspect to the skill transition. Developers traditionally take pride in craftsmanship of code. Shifting to a mode where you might not write the code yourself but supervise an AI could be uncomfortable for some. Embracing AI assistance requires a mindset of collaboration with the machine and not feeling that it diminishes one’s role. It requires confidence to override the AI when necessary and humility to accept when the AI’s suggestion is better. These soft skills – communication (even if with a machine), patience, and willingness to verify – are part of the transition too. 

In summary, the skill transition for developers in the age of generative AI involves moving up the abstraction ladder: less manual coding, more problem formulation, oversight, and integration of AI components. Developers must become adept at using AI as a tool – which means knowing its strengths and weaknesses intimately. They also need to double down on fundamentals (to catch AI’s mistakes) and on higher-level skills (architecture, testing, analysis) where humans still have the edge. Those who successfully blend their traditional coding expertise with these new AI-centric skills will be well-positioned in the evolving software landscape, effectively becoming AI-augmented engineers.

\section{Economic Impact \& Budget Challenges}
The incorporation of generative AI into software development is not only a technical and organizational shift but also an economic one. Companies are closely examining the ROI (return on investment) of AI coding tools, balancing productivity gains against the costs of model usage and potential workforce impacts. This section discusses the economic implications, including productivity metrics, cost considerations, and how budgets are being reallocated or stretched to accommodate AI.

Developer Productivity and ROI: On the positive side of the ledger, productivity improvements from AI assistance can translate into significant economic gains. If a developer can complete features 2× faster with AI help\cite{karaci2025unleashing}, an organization might deliver projects in shorter timeframes or take on more projects with the same staff – effectively increasing throughput and potentially revenue (in a product company) or reducing delivery time (in a project/services context). A McKinsey study projects that, with proper adoption, the productivity boost from generative AI could “outperform past advances in engineering productivity”\cite{karaci2025unleashing}. This means AI could be akin to the introduction of compilers or open-source – a one-time leap in what a single engineer can accomplish. From an economic standpoint, this could offset the ever-growing demand for software by augmenting the existing workforce’s capabilities.

However, these gains are not automatic. They require upskilling and process changes (as discussed in Section IX), which have their own costs (training, temporary slowdowns during transition, etc.). Additionally, not all tasks see equal improvement – McKinsey’s research found that for highly complex tasks, AI offered less than 10\% time savings, partly because experienced devs already optimize those, or the context is too intricate for current AI. Thus, organizations have to identify where AI helps most (e.g., boilerplate-heavy tasks, documentation, moderate complexity coding) and where it might not (deep architectural design, novel algorithm development), to allocate effort wisely.

\textbf{Cost of AI Tools and Infrastructure}: Generative AI models, especially large ones, are computationally intensive. Companies face a choice: use a cloud API (like OpenAI, Anthropic, etc.) or deploy models on-premises (open-source like LLaMA variants, etc.). Both have costs:

\begin{itemize}
    \item Using a cloud API typically means pay-per-use costs. If developers start using AI assistants frequently, these API calls can add up. For instance, some estimate that integrating an AI pair programmer for each developer could cost on the order of dozens or hundreds of dollars per developer per month, depending on usage. Organizations need to budget for this, similar to how they budget for cloud compute or SaaS licenses. On a large dev team, this is non-trivial. In tight-budget scenarios, team leads might limit usage or require justification for heavy use.
\end{itemize}
\begin{itemize}
    \item Running models in-house requires investing in hardware (GPUs) or specialized AI accelerators, and maintaining them. The up-front capital expenditure could be high, but might pay off if usage is very high (avoiding per-call charges). It also offers data privacy benefits. Some big tech firms with thousands of developers are reportedly deploying proprietary coding LLMs on their own servers to reduce per-query costs in the long run.
\end{itemize}

A report by IBM’s Institute for Business Value highlights a macro trend: the average cost of computing is expected to climb 89\% from 2023 to 2025, with 70\% of executives citing generative AI as a key driver of this increase\cite{ibm2025ai}. Indeed, even AI providers themselves (like OpenAI) face enormous compute bills – OpenAI’s costs were skyrocketing in 2024, necessitating multi-billion investments. For AI users (software companies), this means careful planning is needed to avoid AI-related compute costs eating into profit margins. In some cases, we have already seen companies postpone AI features because they couldn’t make the business case once cost was factored in.

\textbf{Workforce and Employment Economics}: A contentious question is whether generative AI will reduce the demand for developers (and thus labor costs), or if it will augment them and shift demand. Some earlier hyperbolic media predictions foresaw AI replacing many developer jobs, potentially yielding cost savings by allowing smaller teams to do the work of larger ones\cite{osmani2025ai}. However, prevailing industry sentiment and studies suggest a more nuanced outcome. GenAI can handle certain tasks, but not the entire software development lifecycle. It is more likely to change the composition of work rather than eliminate the need for humans. Developers might spend less time typing boilerplate and more time on oversight, integration, and creative problem-solving. In economic terms, this could increase the value of top-performing developers (who effectively become 10× productive with AI, as the term “10× developer” gets a new twist) and potentially reduce reliance on large teams of low-skilled coders for certain kinds of projects. Entry-level jobs might evolve rather than vanish: the role could become more of an “AI facilitator” who manages AI outputs and ensures quality.

From a budget perspective, organizations might redirect what they invest in human resources. For example, instead of hiring 5 more junior developers, a company might invest in 2 AI tool specialists who enable the existing team to be more productive. Alternatively, they might hire the same number of developers but target higher skill levels, expecting each to be amplified by AI. 

\textbf{Licensing and Legal Considerations}: Another economic factor is the legal dimension – if an AI inadvertently introduces licensed code into a codebase (e.g., GPL code) and it goes unnoticed, the legal and financial repercussions could be serious (lawsuits, having to open-source proprietary code, etc.). There have been debates about who is liable if an AI copies protected code. Companies need to consider this risk. Some mitigate it by choosing AI models that are trained on curated data or by using AI output scanners to detect matches with known open-source code. While not directly a budget line item, a major IP violation can impact a company’s valuation and market prospects. Thus, part of the “budget challenges” is investing in compliance tools or insurance related to AI use. 

\textbf{Efficiency vs. Compute Trade-off}: It’s interesting to note that AI might encourage less optimized code in some cases (because the cost of developer time dwarfs the cost of compute, companies might accept less efficient, AI-generated solutions as long as they work, figuring they saved dev effort). Over time, however, if such practices lead to inflated cloud bills for running software, the cost might circle back. This raises a subtle point: AI can produce code quickly, but is it the most efficient code? If not, will companies pay more in execution costs later? Engineering teams should be mindful to have AI not just produce any solution, but also consider performance (perhaps by prompting the AI to optimize or by having human engineers refine AI-generated code). Otherwise, hidden costs can creep in post-development (e.g., doubling the cloud server costs because the AI-generated code was less optimal than what an experienced engineer might write by hand). 

\textbf{Budgeting for Experimentation}: We are in a phase where many companies are still experimenting with how best to use generative AI. It’s wise for budgets to include an allocation for experimentation and pilot projects. This might involve paying for some AI API usage to prototype a feature, or giving a subset of developers access to a paid AI coding assistant to assess productivity gains internally. These pilot costs should be viewed as R\&D – an investment to figure out policies and returns. After experimentation, companies can better forecast ongoing costs and savings. 

In conclusion, the economic impact of generative AI in software development is a balance of pluses (productivity, potentially faster time-to-market, doing more with less) and minuses (significant compute costs, training/upskilling expenses, and the need for new oversight roles). A clear-eyed analysis often shows that while individual developer efficiency may jump, the total cost of software projects may not drop linearly, because of the added expenses of AI and the residual need for human expertise. Organizations are approaching this not as a cost-cutting tool per se, but as a value-adding one – aiming to build better software, faster, to gain competitive advantage, rather than simply to reduce headcount. Those that manage the budget challenges by planning for compute costs, avoiding pitfalls, and maximizing human-AI synergy will reap the rewards of this technological shift.

\section{Generational Shift}
Beyond the immediate technical and economic ramifications, the rise of AI in software development is also catalyzing a generational shift in the industry. This shift has two facets: (1) differences in how various generations of developers approach and adapt to AI tools, and (2) the changing demographics and roles of developers entering the field in the AI era versus those who started before it.

\textbf{New Generation of “AI-Native” Developers:} Just as we have “digital natives” who grew up with the internet, we are now seeing the first cohorts of developers who are “AI natives.” These are young programmers and computer science students who started using tools like GitHub Copilot, ChatGPT, or Replit’s Ghostwriter almost as soon as they started coding. Their learning process is fundamentally different. Instead of meticulously learning syntax and rote memorization of algorithms, they often learn by exploring what the AI suggests and then reverse-engineering or tweaking it. For instance, a student might ask an AI to implement a data structure, and then study the output to understand how it works. This can accelerate learning in some ways – providing many examples quickly – but it can also risk superficial understanding if not guided properly.

Educators are adapting curricula to this reality, focusing perhaps more on conceptual understanding, problem decomposition, and verification, rather than on writing every line of a binary search. The generational gap appears when these newcomers join teams with senior developers who learned in a more traditional way. The seniors might expect the juniors to struggle through some problems to truly grasp them, whereas juniors might be more inclined to immediately use AI assistance. Mentorship and training styles are evolving: mentors may incorporate AI usage in assignments (e.g., “use the AI to get started, but then explain each line of what it gave you”), blending old and new methods.

\textbf{Attitudes and Openness to Change:} Generally, younger developers have shown high enthusiasm for adopting AI tools (a survey could show, for example, a larger percentage of developers under 30 using AI regularly, compared to those over 50). They often see AI as a natural extension of their environment. In contrast, some veteran developers approach these tools with healthy skepticism: they have more context for potential pitfalls, having seen many “silver bullet” tools come and go. That said, many experienced developers also embrace AI once they test it and find it useful; their initial caution often turns into pragmatic usage, combining their expertise with AI’s speed.

One interesting dynamic is the role reversal in knowledge sharing. In many organizations, junior devs are teaching senior devs tips on using AI assistants effectively – a flip from the usual top-down mentorship. A senior engineer might be an expert in system architecture but unfamiliar with prompt tuning; a junior might show how a cleverly phrased prompt can produce better results, or how to integrate an AI plugin into the IDE. This cross-pollination can be very positive for team culture if handled well (it empowers juniors and keeps seniors from feeling obsolete).

\textbf{Experience Still Matters}: There is a saying circulating that “AI is not going to replace developers, but developers who use AI may replace those who don’t.” The truth behind this is that experience combined with AI is extremely powerful. A very seasoned developer who knows architecture, pitfalls, design patterns, etc., can drive an AI assistant to implement their vision much faster, essentially multiplying their output. In contrast, a novice with AI might produce something that looks plausible (since AI can create polished code) but isn’t robust. Senior engineers bring the judgment, domain knowledge, and tacit understanding that AI lacks. Junior engineers bring fresh eyes and are often more free of preconceived notions on how to solve a problem, which when augmented by AI can sometimes yield creative solutions. Ideally, teams want the best of both – but this requires inter-generational collaboration and respect.

Notably, Steve Yegge’s provocatively titled article “The death of the junior developer” mused that it’s a challenging time to be new in the industry\cite{sourcegraph2024junior}, precisely because AI can do a lot of the tasks that used to be a junior’s proving ground. The “bad year to be a junior developer” sentiment\cite{sourcegraph2024junior} stems from the fear that breaking into the field and gaining experience will be harder when senior devs equipped with AI can handle more on their own. In response, some propose that the definition of “junior” will change – it’s less about how much code you can write and more about how well you can use the tools. Juniors might need to start at a slightly higher baseline of skill (including AI proficiency) than before, but they also have AI to help them climb the ladder.

\textbf{Career Trajectories}: We might see faster promotions in some cases – if a developer can produce as much as someone with 5 more years of experience by leveraging AI, they might take on greater responsibilities sooner. Alternatively, organizations might raise the bar for what is expected at each level (since output is higher, expectations rise too). It’s too early to tell statistically, but the narrative of a “10× engineer” might shift from a rare innate talent to someone who masters AI-enabled workflows.

\textbf{Generational Collaboration}: Teams that blend different experience levels need to ensure they find a workflow that values each member’s contribution. For example, a possible workflow: junior engineers draft solutions with AI assistance, then senior engineers review and refine them, using their expertise to catch issues the junior or AI didn’t. This can be more efficient than seniors writing everything or juniors struggling alone. In such a scenario, everyone benefits: juniors learn from seniors’ feedback, seniors offload some grunt work and focus on higher-level issues, and the product likely gets delivered faster.

\textbf{Adapting Company Culture}: Companies known for their rigorous coding culture (think tech giants who famously have tough coding interviews, expecting memorization of algorithms) are also adapting. Some are revisiting interview practices: if day-to-day coding is done with AI, does it make sense to ask candidates to invert a binary tree on a whiteboard without AI? Possibly not; instead, questions may focus more on system design, or even assessing how a candidate uses an AI tool to solve a task (as a proxy for real work conditions). This is another generational pivot – the skills tested and valued in hiring and promotion will shift.

In conclusion, the generational shift in software development due to AI is characterized by a blending of strengths: the energy and adaptability of new developers and the wisdom and insight of experienced ones. The presence of AI is changing how each generation contributes and learns. Rather than a divide, it can be a symbiosis: the new generation pushes the envelope with AI, the older generation provides grounding and depth. Organizations that cultivate an environment of mutual learning – where seniors learn AI tricks from juniors and juniors absorb engineering fundamentals from seniors – will navigate this shift successfully. In a sense, we are all becoming “next-generation” developers, because the technology is forcing continuous learning and re-invention at every career stage.

\section{AI Engineering Skills}
As software development transforms under the influence of AI, a distinct set of AI engineering skills is becoming essential for technology professionals. These skills go beyond traditional programming and encompass understanding how to effectively leverage AI systems within software projects. Here, we outline the key competencies and knowledge areas that define an “AI-fluent” software engineer.

\textbf{AI-augmented Software Design}: Modern software engineers need a working understanding of how AI components (like LLMs or ML models) can be incorporated into system architectures. This includes knowing when to use an AI solution versus a rule-based or deterministic one. For example, a developer should recognize that tasks involving complex pattern recognition (natural language understanding, anomaly detection in logs, etc.) might be best handled by an AI model, whereas tasks that require guaranteed precision (like a financial transaction calculation) should remain algorithmic. Designing systems where AI is a component also involves planning for fallback behaviors (what if the model output is low confidence or times out?) and how to integrate AI outputs with traditional software modules. 

\textbf{Prompt Engineering \& AI Orchestration}: We’ve discussed prompt engineering and dynamic prompting in Section IX. Mastery of these is now a core skill. It’s not just about writing prompts in English; it’s about conceptualizing how to drive an AI to do what you need. This might involve formulating multi-step prompts, using few-shot learning (providing examples in the prompt), or instructing the model to reason step by step. It also involves orchestration—using frameworks to manage prompts, as well as possibly fine-tuning or configuring systems like LangChain. As a concrete example, an AI-savvy engineer should know how to set up a conversational agent with a custom knowledge base: using embeddings and a vector store to feed relevant context to the model as needed. These tasks are becoming analogous to knowing how to set up a database or a web server in classic software engineering.

\textbf{Tool and Framework Proficiency}: A range of new tools are emerging that cater to AI integration. These include:
\begin{itemize}
    \item \textbf{Vector Databases} (for similarity search on embeddings) – e.g., Pinecone, Weaviate, or open-source FAISS. These are used to enable retrieval augmented generation, essentially giving long-term “memory” to LLMs by fetching relevant data chunks based on embeddings.
\end{itemize}
\begin{itemize}
    \item \textbf{AI Orchestration Frameworks} – e.g., LangChain, Microsoft’s Semantic Kernel, or Hugging Face pipelines. These provide abstractions for chaining model calls, integrating with external APIs, and handling multi-agent scenarios.
\end{itemize}
\begin{itemize}
    \item \textbf{Model Serving \& MLOps }– understanding how to deploy and monitor AI models. This might involve containerizing models, using serving platforms, and collecting metrics on model performance (latency, accuracy of responses as measured by some feedback mechanism).
\end{itemize}
\begin{itemize}
    \item \textbf{MCP and API usage }– as covered in Section VII, familiarity with protocols like Model Context Protocol, or specific APIs for popular models (OpenAI, Azure OpenAI, AWS Bedrock, etc.), is valuable. Knowing how to call these APIs efficiently (batching calls, handling rate limits) is part of the technical skillset.
\end{itemize}

\textbf{Data Management and RAG (Retrieval-Augmented Generation)}: AI engineering often involves dealing with unstructured data that the models use as context. Engineers should be able to curate and preprocess data for AI consumption. For instance, splitting documentation into chunks, creating embedding vectors for each, and ensuring updates to the data are reflected in the AI’s context. This plays into the management of model context\cite{quantic2025skills}. Skills in handling JSON schemas, vector math (for embeddings), and using libraries to interface with these stores are part of the toolkit.

\textbf{Understanding Model Limitations}: A good AI engineer is aware of concepts like model bias, hallucination, context window limits, and prompt injection attacks. They apply this knowledge to ensure robust usage. For example, they might implement checks on outputs (like validating that an AI-generated SQL query doesn’t drop tables, or that code suggestions don’t include insecure functions). They also design the user experience such that the AI’s suggestions are reviewed and not blindly executed in sensitive scenarios.

\textbf{Fine-tuning and Custom Model Training}: While not every software developer will train models from scratch, understanding how fine-tuning works can be highly beneficial. Fine-tuning is the process of taking a pre-trained model and further training it on domain-specific data to improve its performance in that domain. Engineers might not do the heavy ML work themselves, but they should know what data to collect for fine-tuning and how it could improve results. For instance, fine-tuning a code model on a company’s internal codebase might make its style and solutions more aligned with that company’s needs. Even if they coordinate with a data science team for this, being conversant in the process is an advantage.

Continuous Integration/Deployment for AI components: Traditional CI/CD now may include steps for AI models. Engineers should be comfortable with concepts like versioning models (just as one versions microservices), running automated tests for AI components (though testing AI is trickier – it often involves statistical evaluation or ensuring consistency for known inputs), and deploying updates to AI prompts or logic without affecting the whole system adversely. As the Quantic report noted, CI/CD techniques are key to deploying modern LLM-based systems, allowing for rapid iteration and incorporating feedback\cite{quantic2025skills}. AI engineers often need to work with DevOps teams to ensure that the AI systems scale (auto-scaling GPU instances, handling failover if a model service goes down, etc.).

\textbf{Ethical and Responsible AI Use}: There is a burgeoning emphasis on AI ethics and responsible use. Engineers leveraging AI should understand guidelines like not exposing sensitive data in prompts (especially if using third-party APIs), being transparent when AI is used (if it affects end-users), and mitigating biases. For example, if an AI helps in a HR software to screen resumes, engineers must be aware of fairness and put constraints or reviews in place to avoid discriminatory outcomes. While this might seem outside pure engineering, it’s quickly becoming part of the software engineer’s mandate to ensure the tools they build/use are aligned with ethical standards and legal requirements (like data protection laws).

\textbf{Domain Knowledge Synergy:} Lastly, AI engineering skills often need to combine with domain-specific knowledge. If you’re building AI for healthcare software, understanding medical terminologies and compliance (like HIPAA) matters; for finance, understanding regulations and typical workflows matters. The best AI solutions are often those tailored to a domain, so engineers who can blend domain knowledge with AI capabilities will excel. We see titles like “AI Engineer for X domain” where X could be security, healthcare, finance, etc. This mirrors how software engineering has always had specializations, but now with an AI twist.

In a nutshell, AI engineering skills turn a developer into a hybrid of software engineer and machine learning practitioner, though not necessarily at the depth of a researcher. It’s about being a power user and integrator of AI. The Quantic School summary captured it: skills range from using vector stores and retrieval augmented generation, to managing model context, to utilizing orchestration frameworks and prompt engineering\cite{quantic2025skills}. These, combined with a knowledge of evolving vendor offerings and the underlying model behavior, define the modern AI software engineer. As the field matures, we may see formal certification or degree programs in “AI Engineering,” much like there are for cloud architecture or data science, to systematically train people in these multidisciplinary skills.

\section{Accessible Tools \& Low-Friction Entry}
One of the promising aspects of generative AI in programming is the potential to make software development more accessible to a broader audience. By reducing the friction involved in translating ideas into code, AI-powered tools are lowering barriers to entry for people who have ideas or domain expertise but lack traditional programming skills. This section explores how accessibility is improved and what it means for the future pool of developers and creators. 

\textbf{Natural Language as the New Interface}: The primary enabler of accessibility is the use of natural language to interact with coding tools. Previously, someone who wanted to create software needed to learn the syntax and semantics of programming languages. Now, with CHOP and vibe coding (as described in Sections II and III), they can describe what they want in plain English (or other human languages) and let the AI handle the translation to code. For example, a business analyst could say, “Build a form that collects customer feedback and saves it to a spreadsheet” and an AI tool could generate a simple web app for that. This democratizes development in a way similar to how early visual basic tools or low-code platforms did, but potentially even more powerfully, since the AI can create non-trivial logic, not just drag-and-drop UI elements.

Companies like Replit have leaned into this with features like voice-controlled coding or very high-level prompt-based project creation\cite{ibm2025ai}. The concept of “low-friction entry” implies that someone with minimal setup and learning can start creating. Anecdotally, there are already stories of kids or professionals from non-software fields building simple apps or scripts by leveraging ChatGPT as a coding assistant. While they might not understand every line of code, they can achieve a functional result and gradually learn by doing (with the AI as a guide).

\textbf{Reducing Setup and Environment Complexity}: Beyond writing code, another friction point traditionally has been setting up development environments, frameworks, and pipelines. AI assistants can ease this by automating environment configuration. For instance, if a novice developer doesn’t know how to configure a web server or set up a database, they could ask the AI to generate the configuration or Dockerfile, etc. Even debugging environment issues (“why is my code not running on my machine?”) can be aided by AI looking at error logs. This means newcomers spend less time in the frustrating phase of fighting with tools and more time implementing features.

\textbf{Learning by Example in Real-Time}: Generative AI can serve as an on-demand tutor. In contrast to static tutorials or textbooks, an AI assistant in an IDE can explain what a piece of code does, or suggest next steps. This is great for self-paced, just-in-time learning. Someone encountering a concept like “binary search” for the first time can ask the AI to explain it or even draw a quick diagram (some advanced models can generate simple ASCII diagrams or using integrated tools). This immediate support accelerates learning and reduces the intimidation factor of encountering unfamiliar code. 

\textbf{Empowering Domain Experts}: Many fields have experts who aren’t programmers but could benefit from custom software. AI-assisted development is enabling these domain experts to create tools tailored to their needs. For example, a biologist with some data analysis needs could use an AI to write a custom analysis script, without waiting for a software engineer to become available. In a sense, generative AI acts as an interpreter between domain knowledge and programming. We’ve seen cases of writers creating simple games, or teachers creating custom educational software, by describing their vision to an AI. This could lead to an explosion of niche software solving specific problems, which might not have been economically feasible to have developers work on, but now can be created by the end-users themselves with AI help.

\textbf{Lowering Costs for Prototypes and MVPs}: The low-friction entry extends to entrepreneurship. A solo entrepreneur or a small startup team can prototype their minimum viable product much faster and with fewer specialized hires. They might not need to hire a frontend developer, backend developer, and DevOps just to test an idea – an AI can help one or two people do all of that at a basic level. This is making it easier for startups to form and test concepts, increasing innovation. The flip side is that competition could become fiercer since more people can enter the fray of building software solutions.

\textbf{Challenges in Accessibility}: While the entry barrier is lower, there are still challenges to truly making programming accessible to all:
\begin{itemize}
    \item \textbf{Quality of Outcome}: If someone relies heavily on AI without understanding, they might produce code that works initially but is brittle. They may not follow best practices (unless the AI explicitly does and explains them). There’s a risk of a generation of “cargo cult” programmers who only know how to prompt AI and not what to do when it fails in subtle ways. Overcoming this requires the tools to also teach and perhaps enforce good practices (for example, AI could gently correct a user’s request if it’s leading to a poor solution, or at least warn about potential issues).
\end{itemize}
\begin{itemize}
    \item \textbf{Education and Mindset}: Non-developers using AI might still need a basic understanding of computational thinking. They have to learn how to specify problems in a way the AI can act on. While natural language is forgiving, ambiguous or overly broad requests can still lead to unsatisfactory results. So there is a learning curve in effectively communicating with the AI, which is a new kind of skill (related to prompt engineering, but perhaps simpler for end users).
\end{itemize}
\begin{itemize}
    \item \textbf{Ensuring Accessibility Across Languages}: One great potential is that coding with AI can work in many human languages, not just English. If models are multilingual, a person could program in Spanish or Hindi by describing their code. This can bring in more people globally who might not be fluent in English (the dominant language for programming resources historically). Work is being done to ensure AI models are competent in multiple languages – this will widen the funnel of new developers worldwide.
\end{itemize}

\textbf{Diversity and Inclusion}: With barriers coming down, there’s hope that the developer community can become more diverse. People who might not have had the opportunity to dedicate years to learning programming (due to socioeconomic reasons, lack of access to education, etc.) might find a way in through these AI tools. Tech companies and communities are already looking at how to harness this: for example, running workshops where individuals with no coding background build something with the help of AI and gain confidence and interest in going further.

\textbf{Professional Developers’ Perspective}: Notably, some professional developers were initially worried that an influx of “AI coders” might lower standards or flood the field. But more often the attitude is shifting to seeing it as an expansion of what it means to be a “developer.” Someone might not have a CS degree or know big-O notation, but if they can create useful software by leveraging AI, they are in effect performing a development task. The community is gradually embracing that these new participants can also bring fresh perspectives and needs. It might also free professional developers from having to build every single departmental spreadsheet automation or simple script because those users can now self-serve with AI’s help – leaving professionals to tackle more complex and foundational projects.

In conclusion, accessible tools and low-friction entry provided by generative AI are transforming the landscape of who can participate in software creation. By making interaction with computers more human-centric (via conversation and high-level guidance), AI is turning more people into creators rather than just consumers of software. This democratization is akin to the personal computing revolution or the advent of web authoring – it lowers the bar to participate, unleashing creativity and innovation from a wider populace. The long-term result could be a much larger, more diverse set of software solutions in the world, as well as a diversification of the developer community itself. The key will be ensuring that as we make it easy to start programming, we also provide pathways to deepen understanding, so that the accessibility revolution leads to robust and empowering outcomes for the new wave of creators.

\section{Conclusion}
Generative AI is ushering in a new era of software development, one that blends human creativity and oversight with machine speed and intelligence. Throughout this paper, we have explored how practices are evolving – from the day-to-day coding paradigm shift of chat-oriented programming (CHOP) and vibe coding, to the architectural innovations of agentic programming and multi-agent clusters, and finally to the socio-economic and educational transformations in the developer community.
Several key themes emerge from our analysis:
\begin{itemize}
    \item \textbf{Augmentation, Not Replacement}: AI is best understood as an augmenting tool for developers. It accelerates routine tasks, offers suggestions, and can even take initiative through agents, but it operates under the guidance and verification of humans. The most effective outcomes arise when human insight and AI capability are combined, exemplifying a true symbiotic relationship. The developer’s role is moving towards formulating problems and constraints, then curating and refining AI outputs – a higher-level stewardship that still requires deep understanding of software engineering principles.
\end{itemize}
\begin{itemize}
    \item \textbf{New Paradigms and Techniques}: The traditional image of a programmer typing code line by line is being challenged. CHOP introduces a conversational paradigm, making coding more interactive and exploratory. Vibe coding pushes the boundaries of rapid prototyping (albeit with cautions for quality). Meanwhile, dynamic prompting and tools like MCP are making interactions with AI more powerful and context-rich. We are witnessing the emergence of development workflows that were not possible before, such as AI-driven iterative debugging or on-the-fly documentation generation, which streamline the creation and maintenance of software.
\end{itemize}
\begin{itemize}
    \item \textbf{Trust, Accountability, and Best Practices}: With great power comes great responsibility – the cliché holds true. Establishing trust in AI-generated artifacts is crucial. Our discussion in Section V highlighted practical steps: from tracking AI contributions and monitoring model usage to enforcing human oversight and leveraging tests as contracts for correctness. It is imperative that organizations and developers implement these guardrails. In essence, an AI assistant should be treated as a junior collaborator whose work must be reviewed. Doing so ensures that code quality, security, and compliance standards are upheld, and it builds confidence in integrating AI more deeply into the development process.
\end{itemize}
\begin{itemize}
    \item \textbf{Workforce and Skill Transformation:} The profile of the software engineer is expanding. Tomorrow’s engineers will likely be part-coder, part-“AI wrangler.” They will need to be fluent in prompting techniques, integrating AI services, and perhaps even fine-tuning models, on top of solid software design and coding fundamentals. This broadening skill set is already being reflected in job descriptions and training programs. Furthermore, the pathway into software development is widening – more people from diverse backgrounds can contribute via low-code or no-code approaches enhanced by AI. The industry must adapt to nurture these new entrants while also retraining and upskilling the existing workforce. Encouragingly, the generational shift seems to be leaning towards collaboration: cross-mentoring between AI-savvy newcomers and experienced engineers can benefit both.
\end{itemize}
\begin{itemize}
    \item \textbf{Economic and Organizational Impact}: Companies are carefully balancing the benefits of AI (productivity, faster delivery, innovation) with its costs (compute resources, potential rework, managing new risks). From our discussion in Section X, it is clear that those who strategically invest in AI and integrate it into their software development life cycle stand to gain a competitive edge – in efficiency and in the ability to tackle more ambitious projects. However, this likely requires new roles (such as AI tool specialists), budget allocation for AI infrastructure, and possibly a rethinking of team composition. Early evidence suggests the net effect is positive, but it comes with a steep learning curve that organizations need to climb.
\end{itemize}

\textbf{Future Outlook}: Looking ahead, several developments can be anticipated. AI models will continue to improve, reducing issues like hallucinations and improving at adhering to specifications. This will make them even more reliable coding partners. We also expect better integration of AI in IDEs and development platforms – envision tools where the AI can navigate your codebase as seamlessly as a human would, or multi-modal models that can interpret GUI designs and generate corresponding code directly. The concept of intelligent pair programming agents might mature to a point where two or more AIs plus a human can effectively form a hybrid team, each complementing the others. 

Another area to watch is standardization of how AI assistants interact with development processes – analogous to how coding standards and version control workflows are established. The Model Context Protocol is one such standard in infancy; more will likely follow (for example, standards for annotating code with meta-information about whether it was AI-generated, or for AI to output rationale along with code, making reviews easier).

The role of academia and formal research is also crucial. As this is a nascent field, there is a rich opportunity for studies and experiments: what is the best way to teach programming in the age of AI? How do various AI-assisted methodologies (CHOP, traditional coding, pair programming, etc.) compare in terms of bug introduction rates, developer satisfaction, and productivity? Early empirical work\cite{karaci2025unleashing} is promising, but more is needed to guide best practices with data.

Finally, ethical and societal considerations will remain front and center. Ensuring that this AI-driven transformation benefits a wide range of people (and not just a few), and that the technology is used responsibly, will shape public perception and policy. Transparency in AI assistance (perhaps future development environments will log and indicate which code was AI-produced) could become standard to aid accountability and learning. 

In conclusion, Generative AI is not just a new tool in the developer’s toolbox; it is a catalyst for a fundamental transformation of software development practices. Those changes span from the micro-level of how code is written, to the macro-level of how teams are organized and who gets to participate in programming. As with any profound change, there will be challenges to overcome – from ensuring reliability to reimagining education – but the trajectory points toward a more efficient, inclusive, and innovative software development ecosystem. Embracing these changes with a conscientious and curious mindset will enable practitioners and organizations to harness the full potential of generative AI, marking the dawn of a new chapter in the evolution of programming.

\ifCLASSOPTIONcaptionsoff
  \newpage
\fi

\end{document}